\documentclass[10pt,twocolumn,letterpaper]{article}

\usepackage[noheadfoot,
            left=1in,right=1in,top=1in,bottom=1in,
            columnsep=0.3in
            ]{geometry}
\usepackage{times} 
\usepackage{mathptmx} 
\usepackage[bf,small]{caption} 
\usepackage{url}

\begin{document}
\title{Naughton's Wisconsin Bibliography: A Brief Guide}
\author{Joseph M. Hellerstein\\ Naughton Ph.D.\ Student}
\maketitle 
\thispagestyle{empty}




\begin{quote}
He used to tell me, ``Do what you like to do. It'll probably turn out to be what you do best.''\\
---Wallace Stegner, \emph{Crossing to Safety}

\end{quote}

\section{Introduction}
Over nearly three decades at the University of Wisconsin, Jeff Naughton has left an indelible mark on computer science.  He has been a global leader of the database research field, deepening its core and pushing its boundaries. Many of Naughton's ideas were translated directly into practice in commercial and open-source systems.  But software comes and goes.  In the end, it is the ideas themselves that have had impact, ideas written down in papers.  

Naughton has been a prolific scholar over the last thirty years, with over 175 publications in his bibliography, covering a wide range of topics. This document does not attempt to enumerate or even summarize the wealth of ideas that Naughton has published over the course of his academic career---the task is too daunting.  
Instead, the best this short note aims to do is to serve as a rough map of the territory: something to help other researchers navigate the wide spaces of Naughton's work.

\section{Brief Academic Biography}
Jeffrey F.\ Naughton received his bachelor's degree from the University of Wisconsin in 1982, 
and completed his Ph.D.\ at Stanford in 1987 under the direction of Jeffrey Ullman.  After a two-year stint as an Assistant Professor at Princeton, he returned to the University of Wisconsin where he served on the faculty for 26 years.  During his years at Wisconsin, Naughton supervised numerous research students, served as Chair of the Computer Sciences Department, and was a five-time repeat recipient of the venerable ``Cow Award'' for excellence in classroom teaching.  

His research promise was recognized early with the Presidential Young Investigator award in 1991.  His research success was honored in its fullness multiple times, including the ACM-SIGMOD Test of Time Award in 2004 for Shore object database system~\cite{carey_shoring_1994}, and the 2008 ACM Software Systems Award as a member of the Gamma Parallel Database team.  As overarching recognition of both his contributions and ongoing promise, Naughton received the University of Wisconsin Vilas Associate award for excellence in research in 2000, and was inducted as a Fellow of the ACM in 2002.

As of this writing, Naughton has supervised at least 43 Ph.D.\ students (Table~\ref{phds}). Including the five students he is currently supervising, Naughton has at least 92 Ph.D.\ descendants~\cite{geneaology}.

This bibliography marks a checkpoint in Naughton's career, not an endpoint.  In 2016, Naughton left the University of Wisconsin for a senior post in research and development at Google.  There will undoubtedly be many years of Naughton innovations in his new environment, and hopefully some of those ideas will appear in the scholarly literature as well. 

\section{Bibliographic Overview}
\subsection{Recursive Query Processing (1986-1991)}
Naughton's PhD thesis focused on the optimization of recursive queries in Datalog~\cite{Naughton:1987:ORD:913688}, a topic of significant theoretical interest at the time.  Naughton contributed multiple results in this area at Stanford, Princeton and Wisconsin.  Highlights include techniques begun in his PhD thesis to rewrite queries to avoid recursion entirely when possible (bounded recursions)~\cite{naughton_data_1986,naughton_decidable_1987,naughton_simple_1991}, and techniques to rewrite queries for efficient bottom-up evaluation~\cite{naughton_efficient_1989}.  Sadly, many of Naughton's results in this area are difficult to find online at present; interested scholars are directed to their nearest university library for details.  
  
\subsection{Sampling (1989-1995)}
Sampling and Estimation are recurring themes of Naughton's work throughout his career, with papers as recently as 2016~\cite{wu_sampling-based_2016}.  The heyday of Naughton's sampling work occured in the first half of the 1990's, in the domain of selectivity estimation for query optimization.

Naughton's initial publications on the topic arose during his stint at Princeton, where he worked with celebrated theoretician Richard Lipton. The first paper arose in the domain of recursive queries: estimating the size of transitive closures in order to optimize recursive queries in a cost-based manner~\cite{lipton_estimating_1989}.  This led to subsquent papers on relational database sampling, often with Lipton or Peter J. Haas of IBM Research, covering traditional select/join query selectivity estimation~\cite{lipton_query_1990,haas_fixed-precision_1993,haas_relative_1994,lipton_query_1995,haas_selectivity_1996} as well as distinct value (projection) estimation~\cite{haas_sampling-based_1995}. 

Naughton's expertise on sampling and estimation informed many topics later in his career, discussed below.

\subsection{Object Databases (1991-1997)}
Naughton's work in the database group at Wisconsin is characterized by deep, abiding collaborations on systems projects with his colleagues, notably David DeWitt and Michael Carey.  

First among these efforts was work on Object-Oriented Databases. Particularly influential efforts included the Shore system~\cite{carey_shoring_1994} and the OO7~\cite{carey_oo7_1993,carey_oo7_1993} and Bucky~\cite{carey_bucky_1997} benchmarks.  The OO7 and Shore work are among Naughton's most-cited papers~\cite{scholar}, attesting to the broad interest in Naughton's projects in this area.  At a more technical level, Naughton and his PhD students did deep work in this context on clustering objects in second storage~\cite{tsangaris_stochastic_1991,tsangaris_performance_1992}, database loading for interconnected objects~\cite{wiener_bulk_1994,wiener_oodb_1995}, and garbage collection in secondary storage~\cite{yong_storage_1994}.

\subsection{Parallel Databases (1991-1997)}
Naughton was a participant in Wisconsin's Gamma Parallel Database effort, which is considered one of the landmark research systems in 20th century computing history.  He was also on the team that did follow-on work on Parallel Geo-Spatial data management in the Paradise project~\cite{patel_building_1997}.  Parallelism is another recurring theme in Naughton's work throughout various topics below.

Naughton's contributions in parallel databases focused on improving join processing, including non-equijoins (``band'' joins)~\cite{dewitt_evaluation_1991}, and joins that have to cope with data skew~\cite{dewitt_practical_1992}---the latter being a very common problem in modern Big Data settings.  In these papers, Naughton brought his expertise in database sampling to bear on the runtime execution of queries~\cite{seshadri_sampling_1992}.  Naughton also worked on parallel execution of object traversal~\cite{dewitt_parsets_1994}, as part of a body of work that presaged the popularity of MapReduce-style parallel computation.

\subsection{Aggregation and Data Cubes (1995-2005)}
In addition to joins, Naughton did extensive work on aggregate query processing.  This included early work on adaptive aggregation using sampling~\cite{shatdal_adaptive_1995} in the context of parallel databases.  But some of Naughton's most extensive and well-cited aggregation work was in the area of multidimensional data cubes, including his work on computing the cube~\cite{agarwal_computation_1996,zhao_array-based_1997,zhao_simultaneous_1998,zhao_array-based_1998}, and on working with materialized views of the cube~\cite{deshpande_caching_1998,shukla_materialized_1998,shukla_materialized_2000,luo_locking_2003,luo_locking_2005}.  
Multidimensional data cubes have become a fixture in the practical landscape of data management---they are a standard user interface metaphor in modern Business Intelligence tools and spreadsheets, and are an ongoing area of focus for database vendors as well as open-source database systems.
As of the time of this writing, Naughton's initial work on computing the cube was both his third-most-cited~\cite{agarwal_computation_1996} and seventh-most-cited~\cite{zhao_array-based_1997} papers, with 775 and 527 citations respectively~\cite{scholar}.

\subsection{Document Databases and the Web (1999-2009)}
As the millenium drew to a close, the world became connected via the Internet and the World-Wide Web, and Naughton's work turned to issues in web data management.  These included topics in managing semi-structured document data, and in serving database data online.

Naughton and his students were among the leaders in bridging XML document management and traditional database ideas from the relational era.  The work is broadly applicable to any data model with nested and/or variant structure, including the JSON model used in many currently-popular document databases.

Naughton's work on XML and relational databases was extensive in its scope~\cite{shanmugasundaram_relational_1999,shanmugasundaram_general_2001,halverson_mixed_2003,krishnamurthy_xml-sql_2003,kaushik_integration_2004,krishnamurthy_efficient_2004,krishnamurthy_unraveling_2004,krishnamurthy_xml_2005,doan_case_2009}. It was also extremely influential in both industry and academia.  Naughton's single most widely-cited paper is his 1999 work that opened up this space, laying out the connections and differences between XML and relational databases~\cite{shanmugasundaram_relational_1999}; as of this writing it had over 1500 citations in the literature~\cite{scholar}.  Another topic in this area that has attracted enormous attention is the problem of answering containment queries~\cite{ramasamy_set_2000,zhang_supporting_2001}; the second of these papers is Naughton's second-most-cited result, with over 1000 citations as of the time of writing~\cite{scholar}.

Naughton also returned to his roots in this domain, revisiting problems such as selectivity estimation~\cite{aboulnaga_estimating_2001,aboulnaga_building_2003} and recursive query processing~\cite{krishnamurthy_recursive_2004} in the context of XML.  In later years, Naughton returned to the topic of document data in the guise of ``sparse'' relational datasets, which can be viewed either as relations with many nulls, or key-value maps~\cite{beckmann_extending_2006,chu_case_2007}

A related topic at the turn of the millenium was the integration of databases with web servers.  In this domain, Naughton worked on a series of papers regarding web caching for database-backed websites~\cite{luo_active_2000,luo_form-based_2001,luo_middle-tier_2002,luo_form-based_2008}.

\subsection{Streaming, Progress and Online Query Processing (2001-2014)}
Naughton was an ongoing contributor to improving the user experience for long-running queries---an abiding issue in large-scale analytics.  Naughton's work included an ongoing effort into progress indicators for long-running queries~\cite{luo_toward_2004,luo_increasing_2005,luo_multi-query_2006,li_toward_2013,wu_towards_2013,wu_predicting_2013,wu_uncertainty_2014}, as well as providing online results for those queries while they are in progress~\cite{luo_scalable_2002,luo_non-blocking_2002, lang_partial_2014}.  In both cases, progress and answers often need to be estimated, again exercising Naughton's expertise in database sampling and estimation.  

In a related vein, Naughton contributed fundamental work on processing continuous queries over Data Streams~\cite{viglas_rate-based_2002,kang_evaluating_2003,ayad_approximating_2006,wang_utility-maximizing_2013}---here too, results have to be produced before data is fully consumed.  The setting for much of his work was the Niagara Internet Query System~\cite{naughton_niagara_2001}, a vision of streaming XML documents that combined challenges in XML document processing with challenges in stream processing, adaptive query execution, data integration and text search.

\subsection{Privacy in Databases (2009-2013)}
In his last decade at Wisconsin, Naughton became interested in the topic of data privacy, with a particular focus on anonymization methods for query processing.  In many of Naughton's papers in this area there were connections to topics where he had done pioneering prior work previously, including connections between anonymization and spatial indexing~\cite{iwuchukwu_k-anonymization_2007}, as well as anonymization of set-valued attributes~\cite{he_anonymization_2009}, streaming events~\cite{he_complexity_2011}, range predicates~\cite{zeng_optimal_2013} and recursive queries~\cite{zeng_differentially_2013}.  Naughton's privacy work also including dynamic anonymization~\cite{he_preventing_2011} and anonymization of frequent itemset algorithms for data mining~\cite{zeng_differentially_2012}.

\subsection{Data Provenance and Information Extraction (2006-2012)}
Another topic that Naughton explored in his final decade at Wisconsin was that of extracting data from source systems, and reasoning about the provenance (lineage) of data.  This includes work both in the context of logs from job scheduling systems like Wisconsin's long-running Condor project~\cite{reilly_exploring_2006,reilly_transparently_2009}, as well as automated information extraction from text~\cite{huang_k-relevance:_2007,huang_provenance_2008,reilly_instrumenting_2012} where Naughton also did core research~\cite{kang_schema_2003,chu_relational_2007,shen_declarative_2007,doan_information_2008,kang_schema_2008}.

\subsection{Text Search in Databases (2009-2015)}
Related to his work on XML as well as Information Extraction, Naughton and his students worked on various problems in searching and combining textual data in databases.  This includes work on combining keyword search results with forms~\cite{chu_combining_2009,baid_toward_2010}, approximate string membership~\cite{sun_token_2011,sun_approximate_2012}, and debugging of ``why not'' provenance in keyword search over databases~\cite{baid_debugging_2015}.

\subsection{Indexing (1995-2014)}
A cross-cutting topic in Naughton's work is the development and use of index structures, from generalized search trees~\cite{hellerstein_generalized_1995} to document store indexes~\cite{kaushik_covering_2002,kaushik_updates_2002} to text queries~\cite{kaushik_integration_2004,rae_-rdbms_2014}.

\subsection{And So Much More}
Jeff Naughton is \emph{sui generis}: beyond category. So it is not surprising that the categories above do not cover his work.  Given his devotion to his many Ph.D.\ students, the best overview of Naughton's work may be the topics of his students' dissertations in Table~\ref{phds}.  To both Jeff and his students, I apologize both for the work I misclassified above, and the work I neglected to classify entirely.

\section{A Personal Note}
Jeff Naughton's bibliography and papers, impressive as they are, present only a narrow picture of the man.  I consider myself lucky to have studied with him at Wisconsin.  As Jeff's student both during and after my Ph.D., I learned many things beyond computer science.  I learned how to shake off disappointment and failure, and turn them into research results; I will be forever grateful for his confidence in me and his gentle guidance through difficult times.  I saw how humor can smooth the ups and downs of learning. I was given patient lessons in balancing ambition and grace, from a role model who coupled a characteristically midwestern humility to deep insight, steady confidence and a wicked sense of humor. Perhaps most significantly
I got to see---first with puzzlement and later with admiration---how a first-rate scholar can protect his time, put family first, and raise delightful children. As the years have passed, I've had to find my own path through similar issues, and I've been grateful to have seen the trails Jeff blazed.  I don't try to follow Jeff directly; he is unique.  But he has been a guidepost for me in many issues at the juncture of scholarship, drive, and fulfillment. For all that I am grateful. 

Like Jeff, I have lifelong ties to Madison; probably this makes me more wistful about his departure from the UW than I might be otherwise.  To me, Jeff has many of the qualities that represent the best of Wisconsin character: wit, wisdom and friendly modesty. His departure will leave a hole at the heart of the computer sciences department. I know Jeff will bring all those qualities and more to his new career in industry. I hope the people of both Wisconsin and the data management community continue to benefit from Jeff's brilliance and character for many years to come.

\begin{quote}
And so, by circuitous and unpredictable routes, we converge toward midcontinent and meet in Madison, and are at once \linebreak drawn together.\\
---Wallace Stegner, \emph{Crossing to Safety}
\end{quote}

\begin{figure*}[th]
\begin{tabular}{l|l|p{4in}}
Name & Ph.D. & Dissertation Title\\ \hline
S. Seshadri  & 1992 & Probabilistic Methods in Query Processing\\
Emmanuel Tsangaris  & 1992 & Principles of Static Clustering for Object Oriented Databases\\
Joseph M.\ Hellerstein  & 1995 & Optimization and Execution Techniques for Queries with Expensive Methods\\
Bradley Rubin  & 1995 & Information Retrieval Using a Combined Object-Oriented Database/File System Paradigm\\
Janet Wiener  & 1995 & Algorithms for Loading Object-Oriented Databases\\
Srikant Ramakrishnan  & 1996 & Fast Algorithms for Mining Association Rules and Sequential Patterns\\
Ambuj Shatdal  & 1996 & Architectural Considerations for Parallel Query Evaluation Algorithms\\
Shivakumar Venkataraman  & 1996 & Global Memory Management for Multi-Server Database Systems\\
Yihong Zhao  & 1998 & Performance Issues of Multi-Dimensional Data Analysis\\
Prasad Deshpande  & 1999 & Efficient Database Support for OLAP Queries\\
Amit Shukla  & 1999 & Materialized View Selection for Multidimensional Datasets\\
Karthikeyan Ramasamy  & 2001 & Efficient Storage and Query Processing of Set-Valued Attributes\\
Jayavel Shanmugasundaram  & 2001 & Bridging Relational Technology and XML\\
Ashraf Aboulnaga  & 2002 & Cost Estimation Techniques for Database Systems\\
Qiong Luo  & 2002 & Caching for Web-Based Database Applications\\
Chun Zhang  & 2002 & Relational Databases for XML Indexing\\
Jaewoo Kang  & 2003 & Toward the Scalable Integration of Internet Information Sources \\
Raghav Kaushik  & 2003 & Graph Summarization for Path Indexing in Graph-Structured Data\\
Stratis Viglas  & 2003 & Novel Query Optimization and Evaluation Techniques\\
Rajasekar Krishnamurthy  & 2004 & XML-to-SQL Query Translation\\
Gang Luo  & 2004 & Techniques for Operational Data Warehousing\\
Ahmed Ayad  & 2006 & Optimization and Approximation Techniques for Data Streaming Queries\\
Jennifer Beckmann  & 2006 & Relational Database Management System Support for Sparse Data Sets\\
Alan Halverson  & 2006 & Storage and Query Processing Optimizations for Hierarchically-Organized Data\\
Tochukwu Iwuchukwu  & 2007 & Anonymization Techniques for Large and Dynamic Data Sets\\
Ameet Kini  & 2007 & Supporting Match Joins in Relational Database Management Systems\\
Eric Chu  & 2008 & Sparse Relational Data Sets: Issues and an Application\\
Jiansheng Huang  & 2008 & On Interpreting and Debugging Results of Database Queries over Imprecise Data\\
Lakshmikant Shrinivas  & 2008 & Applications of Data Mining to Cluster Scheduling and Failure Diagnosis\\
Christine Reilly  & 2010 & Transparent Gathering of Provenance During Program Execution\\
Akanksha Baid  & 2011 & Toward Scalable Keyword Search over Relational Data\\
Yeye He  & 2012 & Privacy Preserving Data Publishing and Analysis\\
Chong Sun  & 2012 & Multi-Filter String Matching and Human-Centric Entity Matching for Information Extraction\\
Khai Tran  & 2013 & Realizing Parallelism in OLTP Workloads\\
Chen Zeng  & 2013 & On Differentially Private Mechanisms for Count-Range Queries and their Applications\\
Yueh-Hsuan Chiang  & 2014 & Towards Large-Scale Temporal Entity Matching\\
Jiexing Li  & 2014 & Performance Prediction and Resource Bricolage for Database Systems\\
Ian Rae	& 2014 &	From Index Nested Loops to ZigZag Merge: An Experimental Analysis of Skipping Join Algorithms\\
Wentao Wu & * &  \\
Arun Kumar & * & \\
Fatemah Panahi & * & \\
Xi Wu & * & \\
Bruhathi Sundarmurthy & * &\\
\hline
\end{tabular}

\noindent
* \emph{Ph.D. expected.}
\caption{Jeff Naughton's Ph.D.\ Students, 2016}
\label{phds}
\end{figure*}

\nocite{luo_non-blocking_2002,ramasamy_set_2000,sun_approximate_2012,dewitt_practical_1992,shukla_storage_1996,chiang_tracking_2014,shen_declarative_2007,naughton_simple_1991,naughton_sigmod2000_2000,carey_oo7_1993,lipton_estimating_1989,luo_transaction_2008,deshpande_cubing_1997,aboulnaga_building_2003,reilly_transparently_2009,zeng_differentially_2012,naughton_one-sided_1987,reilly_exploring_2006,haas_fixed-precision_1993,kang_evaluating_2003,shankar_integrating_2005,carey_status_1994,reilly_instrumenting_2012,abadi_beckman_2014,kaushik_updates_2002,lang_partial_2014,estan_end-biased_2006,luo_middle-tier_2002,deshpande_aggregate_2000,aboulnaga_generating_2001,krishnamurthy_recursive_2004,halverson_mixed_2003,krishnamurthy_difficulty_2003,hellerstein_query_1996,doan_case_2009,wu_uncertainty_2014,naughton_guest_1998,doan_case_2009,li_low-latency_1994,do_turbocharging_2011,carey_oo7_1993,chi_distribution-based_2013,haas_relative_1994,zhao_array-based_1998,li_gslpi:_2012,xiong_software-defined_2014,aboulnaga_estimating_2001,huang_k-relevance:_2007,dewitt_evaluation_1991,naughton_redundancy_1986,kang_schema_2003,dewitt_clustera:_2008,he_anonymization_2009,luo_scalable_2002,naughton_how_1994,naughton_niagara_2001,wang_utility-maximizing_2013,bernstein_asilomar_1998,morris_yawn!_1987,lipton_practical_1990,tran_transactional_2010,viglas_maximizing_2003,he_complexity_2011,li_toward_2013,naughton_minimizing_1989,naughton_efficient_1989,luo_active_2000,aboulnaga_accurate_2000,seshadri_sampling_1992,kaushik_integration_2004,abiteboul_lowell_2003,li_real-time_1990,haddad_counting_1991,lipton_clocked_1993,krishnamurthy_xml-sql_2003,shukla_materialized_2000,abadi_beckman_2016,he_load_2014,dewitt_we_2013,luo_increasing_2005,kaushik_synopses_2005,kini_database_2007,kumar_demonstration_2015,naughton_dbms:_2010,luo_locking_2005,wu_sampling-based_2016,baid_toward_2010,krishnamurthy_xml_2005,baid_toward_2010,he_load_2013,wu_predicting_2013,abiteboul_lowell_2005,deshpande_caching_1998,shatdal_cache_1994,krishnamurthy_unraveling_2004,naughton_bottom-up_1991,seshadri_expected_1995,kumar_learning_2015,zhao_simultaneous_1998,naughton_decidable_1987,krishnamurthy_efficient_2004,doan_information_2008,lipton_efficient_1993,chen_design_2002,zhao_array-based_1997,li_efficient_1991,haas_sampling-based_1995,chai_efficiently_2009,kang_schema_2008,chiang_modeling_2014,venkataraman_memory_1997,li_checkpointing_1991,luo_comparison_2003,kini_database_2006,shrinivas_issues_2008,he_preventing_2011,seshadri_expected_1991,naughton_argument_1995,li_resource_2014,iwuchukwu_k-anonymization_2007,viglas_rate-based_2002,huang_provenance_2008,li_multiprocessor_1988,huang_trac:_2006,dewitt_parallel_1991,naughton_one-sided_1991,zeng_optimal_2013,luo_locking_2003,lang_energy_2009,sun_token_2011,shanmugasundaram_general_2001,luo_active_2000,shukla_materialized_1998,haddad_counting_1988,chu_combining_2009,shanmugasundaram_architecting_2000,luo_form-based_2001,dewitt_parallelising_1996,luo_transaction_2010,subramanian_impact_2010,naughton_data_1986,agarwal_computation_1996,tran_jecb:_2014,yong_storage_1994,shatdal_using_1993,wiener_bulk_1994,ross_reminiscences_2003,bernstein_asilomar_1998,haas_selectivity_1996,rae_-rdbms_2014,naughton_compiling_1988,naughton_estimating_1990,tsangaris_performance_1992,dewitt_nested_1993,wu_revisiting_2015,wiener_oodb_1995,venkataraman_impact_1995,chaudhuri_relational_2003,ayad_approximating_2006,luo_toward_2004,patel_building_1997,shanmugasundaram_relational_1999,hellerstein_generalized_1995,naughton_one-sided_1986,lipton_query_1995,tsangaris_stochastic_1991,luo_multi-query_2006,carey_bucky_1997,zeng_differentially_2013,shatdal_adaptive_1995,naughton_argument_1989,iwuchukwu_k-anonymization_2007,ayad_static_2004,wu_uncertainty_2014,luo_transaction_2006,wu_towards_2013,chu_relational_2007,cai_complexity_2001,luo_form-based_2008,kaushik_covering_2002,kaushik_integration_2004,luo_transaction_2008,zhang_supporting_2001,chu_case_2007,venkataraman_remote_1998,carey_shoring_1994,naughton_how_1990,gokhale_corleone:_2014,dewitt_parsets_1994,lipton_query_1990,srivastava_space_1995,shanmugasundaram_architecting_2000,baid_debugging_2015,sudarshan_space_1991,naughton_data_1989,beckmann_extending_2006}
\bibliographystyle{habbrvyr}
\bibliography{naughton}

\end{document}